\documentclass{aastex}
\usepackage[utf8]{inputenc}
\usepackage{spr-astr-addons}
\usepackage{url}
\usepackage{graphicx}
\usepackage{natbib}

\RequirePackage{color}

\newcommand{\mfar}{MF$_{\text{ar}}$}
\newcommand{\mfqr}{MF$_{\text{qr}}$}

\begin{document}

\title{Modulations of the Surface Magnetic Field on the Intra-cycle Variability of Total Solar Irradiance}
\shortauthors{J. C. Xu et al.}
\author{J. C. Xu\altaffilmark{1,2,3}} \and 
\author{D. F. Kong\altaffilmark{1}} \and 
\author{F. Y. Li\altaffilmark{1,4}}

\altaffiltext{1}{Yunnan Observatories, Chinese Academy of Sciences, Kunming 650011, China}
\altaffiltext{2}{State Key Laboratory of Space Weather, Chinese Academy of Sciences, Beijing 100190, China}
\altaffiltext{3}{Key Laboratory of Solar Activity, National Astronomical Observatories, Chinese Academy of Sciences, Beijing 100012, China}
\altaffiltext{4}{University of Chinese Academy of Sciences, Beijing 100049, China}
\email{jcxu@ynao.ac.cn}

\begin{abstract}
Solar photospheric magnetic field plays a dominant role in the variability of total solar irradiance (TSI). The modulation of magnetic flux at six specific ranges on TSI is characterized for the first time.
The daily flux values of magnetic field at four ranges are extracted from MDI/{\sl SOHO}, together with daily flux of active regions (\mfar) and quiet regions (\mfqr); the first four ranges (MF$_{1-4}$) are: 1.5--2.9, 2.9--32.0, 32.0--42.7, and 42.7--380.1 ($\times 10^{18}$ Mx per element), respectively. 
Cross-correlograms show that MF$_4$, \mfqr, and \mfar\ are positively correlated with TSI, while MF$_2$ is negatively correlated with TSI; the correlations between MF$_1$, MF$_3$ and TSI are insignificant. The bootstrapping tests confirm that the impact of MF$_4$ on TSI is more significant than that of {\mfar} and {\mfqr}, and {\mfar} leads TSI by one rotational period. 
By extracting the rotational variations in the MFs and TSI, the modulations of the former on the latter at the solar rotational timescale are clearly illustrated and compared during solar maximum and minimum times, respectively. Comparison of the relative amplitudes of the long-term variation show that TSI is in good agreement with the variation of MF$_4$ and {\mfar}; besides, MF$_2$ is in antiphase with TSI, and it lags the latter by about 1.5 years.
\end{abstract}

\section{INTRODUCTION}
\label{sect:intro}

Total solar irradiance (TSI) is the entire radiative power per unit area measured above the Earth’s atmosphere and normalized to the mean Sun-Earth distance of one AU, given in SI units of Watts per square meter (W m$^{-2}$). TSI is  the major external source of energy input into the Earth's climate system; therefore, besides other mechanisms such as anthropogenic greenhouse effect and ozone variations, the variability of TSI is considered as one of the key and most obvious factors that lead to the climate change during the last century (see reviews, e.g.,~\citealt{haigh_sun_2007, gray_solar_2010, solanki_solar_2013}, and references therein). Additionally, TSI is a fundamental parameter of the Sun itself as a star. 
The space-borne observations since late 1978 reveal that TSI is varying at all of the timescales at which it has been measured, i.e., from minutes to decades~\citep{lean_variations_1991,  frohlich_solar_2004, unruh_spectral_2008, krivova_towards_2011, li_why_2012, frohlich_total_2013, wehrli_correlation_2013}. 

It is widely acknowledged that the variation of  TSI is intimately related with the solar surface magnetic field. Examinations with the measurements of TSI during the last four decades show that the variation of surface magnetism dominates the variation of TSI at timescales from a day to the Schwabe cycle~\citep{domingo_solar_2009}. It is believed that the large-scale solar magnetic field is generated by a global large-scale dynamo in the convection zone; this type of dynamo, which is best shown by sunspots and the solar cycle, is caused by the $\alpha$ effect together with the $\Omega$ effect~\citep{charbonneau_solar_2014}. With the development of observational techniques, it is found that small-scale magnetic elements are everywhere at anytime on the solar surface. These small-scale magnetic fields are speculated to be caused by local small-scale dynamo processes~\citep{karak_is_2016}. However, it is also possible that these small-scale magnetic fields may mainly come from a higher latitude or emerge from the subsurface shear layer~\citep{xu_rotation_2016}.

Based on our understandings of the TSI variation, various models are built to reconstruct TSI, so as to validate our understandings and extend the TSI data set for the sake of many other scientific topics. 
Proxies of the solar surface magnetic field such as the sunspot numbers, the photometric sunspot index, the Mg II index , the F10.7 radio flux, and the Lyman-$\alpha$ flux are commonly used to reconstruct TSI with proxies models. Usually, these indices represent either the darkening effect of sunspots or the brightening effect of faculae and networks \citep{hudson_observed_1988, frohlich_solar_2004}. The 4-component proxy model in \cite{frohlich_total_2012} is proved to be capable of explainning 84.7\% of the TSI variance in the Physikalisch-Meteorologisches Observatorium Davos (PMOD) composite data during 1978-2011. 
A prominent empirical proxy model is the Naval Research Laboratory Total Solar Irradiance (NRLTSI) \citep{lean_evolution_2000, coddington_solar_2015}, and the correlation between the model and the Total Irradiance Monitor/Solar Radiation and Climate Experiment (TIM/{\sl SORCE}) observation is 0.96.
Another type of successful model is the semi-empirical model. Various authors have used solar magnetograms and photometric images to reconstruct the solar total and spectral irradiance~\citep[see, e.g.,][]{foukal_influence_1986, fligge_model_1998, fligge_modelling_2000, krivova_reconstruction_2003, krivova_reconstruction_2006, ball_reconstruction_2012, yeo_reconstruction_2014}. 
\cite{ball_reconstruction_2012} found that the reconstructed TSI can account for 92\% of the variations in the PMOD~\citep{frohlich_suns_1998}, and over 96\% of that in solar cycle 23; the reconstruction is carried out with magnetograms and continuum images from the Michelson Doppler Imager on board the Solar and Heliospheric Observatory (MDI/{\sl SOHO}) and the Kitt Peak Observatory, with an assumption that the TSI variation is solely caused by the changes of the photospheric magnetic field. 
\cite{dasi-espuig_reconstruction_2016} reconstructed TSI back to 1700 with simulated magnetogram based on a surface flux transport model.
In addition, as for the various long-term reconstructions of TSI, \cite{soon_re-evaluating_2015} conclude that they can be categorized into two main groups according to whether they show a substantial increase since the Maunder Minimum, namely the high- and low solar variability family.

Despite the great progress in reconstructing the TSI based on the surface magnetic field or its proxies, the relation between the small-scale magnetic field and TSI  is not yet clear. 
\cite{li_why_2012} examined the variation of TSI on the inter-solar-cycle scale, and concluded that it is constituted by three parts: a rotational variation part which is caused by large magnetic structures, an  annual variation part which might be caused by the annual change of Earth's helio-latitude, and an inter-solar-cycle part which is caused by the network magnetic elements in the range (42.7--380.1) $\times$ 10$^{18}$ Mx per element.
Using the Magnetic Plage Strength Index and Mount Wilson Sunspot Index which roughly represent the weak and strong solar magnetic field activity, \cite{xiang_what_2015} studied the relation between magnetic field and TSI, and also found that the weak magnetic field dominates the inter-solar-cycle variation.

In this study, we characterize the impact of magnetic flux of specific ranges on the intra-cycle variability of TSI with magnetic flux data extracted from the magnetograms of MDI/{\sl SOHO}. 
In Section~\ref{sect:data} the data used in this research are introduced briefly. The methods, analyses, results, and discussions are presented in Section~\ref{sect:MRD}. In the last section, we conclude the results.

\section{Data}
\label{sect:data}

The data used in this research include daily values of six categories of magnetic flux (MF) and one  TSI composite data.

\subsection{Magnetic Flux}
Since the discovery of magnetism in sunspots~\citep{hale_probable_1908}, it was found out soon that magnetic field can be detected everywhere in the solar atmosphere~\citep{stenflo_history_2015}. However, at present, only magnetic fields in the photosphere can be measured precisely with optical instruments. 
Within the ability of the current resolutions of telescopes, besides obviously recognizable active regions with strong magnetic fields, magnetic elements are found to be sprinkled all over the photosphere~\citep{sheeley_measurements_1966, harvey_solar_1971}. The distinguishable small-scale magnetic elements are mainly composed of network elements, intra-network elements, and ephemeral regions. 
\cite{jin_suns_2011} decomposed the magnetograms of  MDI/{\sl SOHO} (daily) and extracted millions of small-scale elements; these magnetic elements were grouped into several categories according to their flux quantities (Mx per element) and their relations with the sunspot numbers. They obtained flux data sets of magnetic elements in four different ranges, together with the total flux of active regions and quiet regions. The full-disk magnetograms of the MDI/{\sl SOHO} that were used for extraction cover the time interval from 1996 September to 2010 February, which is about 13.5 years and includes the whole solar cycle 23. 

In the process of decomposition, one 5-minute averaged magnetogram is extracted on each day, and it is smoothed to further reduce the noise level which is determined to be 6 Mx cm$^{-2}$.
It is assumed that the observed line-of-sight magnetic field is a projection of the intrinsic magnetic field whose direction is normal to the solar surface, and thus the MF in each pixel is corrected according to its location on the magnetogram.
 The MF of pixels with heliocentric angles larger than 60$^{\circ}$ is set to 0 due to the scarce magnetic signals.
The threshold of the edge for active regions is 15 Mx cm$^{-2}$, and only islands within a heliocentric angle of 60$^{\circ}$ with an area larger than 9$\times9$ pixels, or islands smaller than $9\times9$ pixels within 60$^{\circ}$ but larger than $9\times9$ pixels within 70$^{\circ}$, are considered as active regions.
After the active regions are excluded, magnetic concentrations with more than 10 pixels in size in the quiet magnetograms are recognized as network magnetic elements \citep{hagenaar_properties_2003, jin_suns_2011}.
In total, over 13 million magnetic elements are identified, which include active features and network magnetic elements.
Observationally, these small-scale network elements are mainly from the decayed fragmentations of active regions, the flux emergence in the form of ephemeral regions, the coalescence of intra-network flux, and products of the interaction among different sources of magnetic flux~\citep{jin_suns_2011, jin_variation_2014}.

\begin{figure}[tbp]
 	\centerline{\includegraphics[width=0.48\textwidth,clip=]{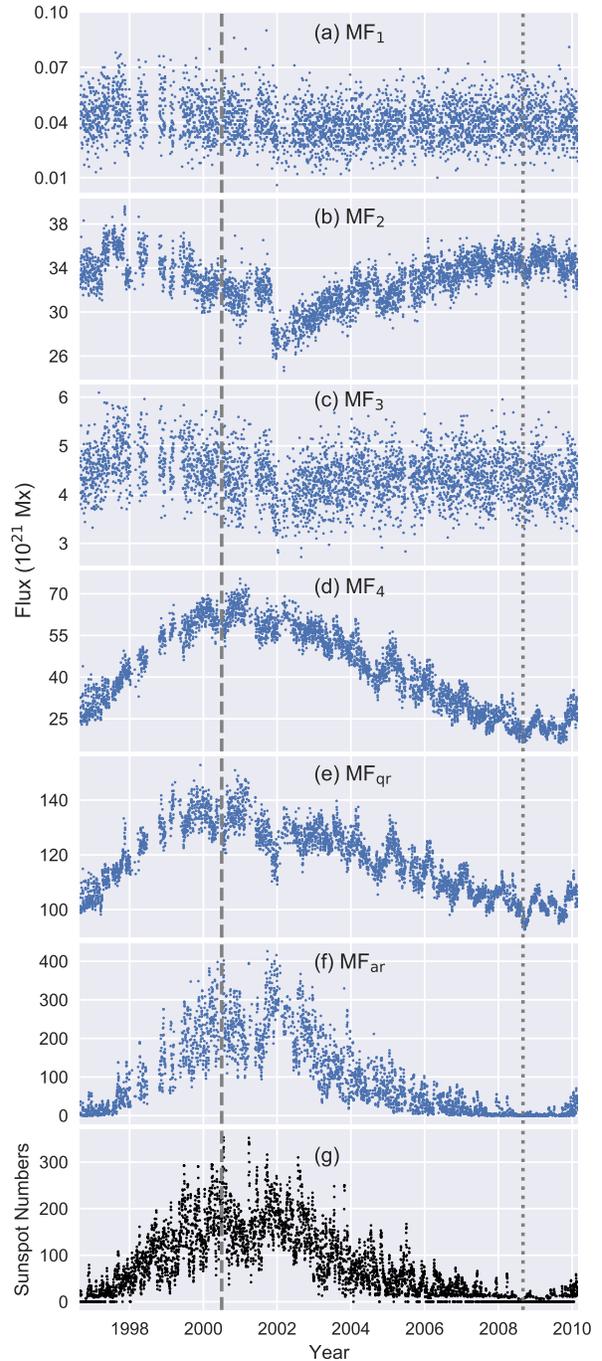}}
	\caption{Daily values of the solar photospheric magnetic flux at four ranges: (a) (1.5--2.9)$\times10^{18}$, (b) (2.9--32.0)$\times10^{18}$, (c) (3.20--4.27)$\times10^{19}$, (d) (4.27--38.01)$\times10^{19}$ Mx per element, as well as the daily values of total flux of  quiet regions (e) and active regions (f). As a comparison, the sunspot numbers are also shown in the bottom panel (g). The time interval is from 1996 September to 2010 February, which covers the solar cycle 23. The vertical gray dashed (dotted) line indicates the time of solar maximum (minimum). These data sets are extracted by~\cite{jin_suns_2011} from daily 5-minute averaged magnetograms from the MDI/{\sl SOHO}. \label{fig_data}}
\end{figure}

These magnetic elements are sorted according to their flux and their correlation with the sunspot numbers. They can be categorized into four flux ranges, together with the total flux of active regions (\mfar) and quiet regions (\mfqr). 
The values for the first four ranges (MF$_{1-4}$) are: 1.5--2.9, 2.9--32.0, 32.0--42.7, and 42.7--380.1 ($\times 10^{18}$ Mx per element). All of the data sets are plotted in Figure~\ref{fig_data}. 
Note that the four ranges are meant for individual magnetic elements, while the data sets shown in Figure~\ref{fig_data} are the daily total flux at each range.
As indicated in \cite{jin_suns_2011},  the MF$_{1-4}$ are the no-correlation elements, anti-phase elements, transition elements, and in-phase elements, respectively, depending on their relation with the sunspot numbers.
The fluxes in active regions and quiet regions, i.e., \mfar\ and \mfqr, are in-phase with the solar activity cycle.

\subsection{TSI}
\begin{figure}[tbp]
 	\centerline{\includegraphics[width=0.48\textwidth,clip=]{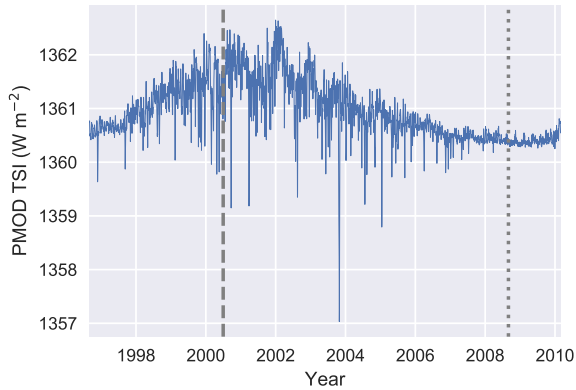}}
	\caption{Daily TSI records of the PMOD composite data from  1996 September to  2010 February. The vertical gray dashed (dotted) line indicates the time of solar maximum (minimum). \label{tsi_data}}
\end{figure}

The precise TSI measurements with space-borne instruments began at late 1978. A complete list and the detailed descriptions of the instruments (there are nearly ten instruments) can be found in, e.g., \citet{frohlich_total_2012}. Due to the limited lifetime of an individual instrument, various observations have to be composed into a continuous sequence, which requires laborious efforts. There are mainly three composite data presented, namely the Active Radio Irradiance Monitor (ACRIM)~\citep{willson_total_1997, willson_secular_2003}, the PMOD~\citep{frohlich_suns_1998, frohlich_solar_2006}, and the Royal Meteorological Institute of Belgium (IRMB)~\citep{dewitte_measurement_2004}. 
The differences between the three composite data sets are discussed in~\cite{frohlich_total_2012} and~\cite{zacharias_independent_2014}.
There are still debates about which one represents the real variation of TSI (especially the intra-cycle variation); some people believe PMOD is more reliable~\citep{gray_solar_2010}, while others prefer ACRIM~\citep{soon_re-evaluating_2015}. 
The new composite proposed by~\cite{dudok_de_wit_methodology_2017} is probably helpful in resolving this issue.
In this study, we will use the PMOD composite data as a representative of TSI, and it is shown in Figure~\ref{tsi_data}. The currently accepted average value of TSI is $1,360.8\pm0.5$ W m$^{-2}$~\citep{kopp_new_2011}, and its long-term variation is obviously inphase with the $\sim$11-year solar activity cycle.

\section{Methods, Results, and Discussions}
\label{sect:MRD}

\subsection{Correlogram and Bootstrapping}
\begin{figure}[tbp]
 	\centerline{\includegraphics[width=0.48\textwidth,clip=]{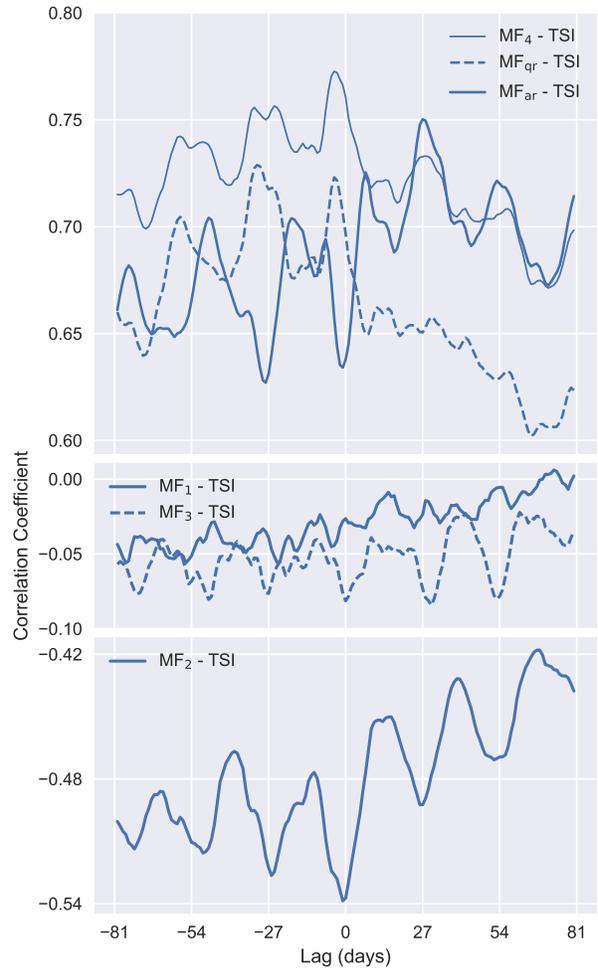}}
	\caption{Cross-correlograms between the six types of magnetic flux and the PMOD TSI composite data (upper panel: MF$_4$ (thin), \mfqr (dashed), \mfar (thick); middle panel: MF$_1$ (solid), MF$_3$ (dashed); bottom panel: MF$_2$). A cross-correlogram is a plot of correlation coefficients versus time lags.
	\label{fig_xc}}
\end{figure}

The lagged cross-correlation analysis, i.e., the well-known Pearson correlation between two lagged time series, is used here to investigate the relation between the MFs and TSI. This method is widely used for determining the mutual correlation and phase relation between two time series~\citep[e.g.][]{peterson_uncertainties_1998, li_phase_2016}. The detailed calculation method is described in~\citet{xu_rotation_2016}. An advantage of this method is that there is no need to take care of data gaps which are simply excluded from the analysis.
The length of the MF data sets is 4929 days (from 1996 September 1 to 2010 Feburary 28), and on 1139 days there are no data.
In the calculation of the cross-correlation analysis, the missing data are just excluded.
The cross-correlograms between the six types of MF and TSI are shown in Figure~\ref{fig_xc}. A cross-correlogram is a plot of cross-correlation coefficients against time lags. 

As can be seen in  Figure~\ref{fig_xc}, the correlation coefficient (abbreviated as `CC' throughout) indicates that MF$_4$, \mfqr, and \mfar\ (the upper panel) are positively correlated with TSI, while MF$_2$ (the bottom panel) is anticorrelated with TSI. The rest of the two types of flux, i.e., MF$_1$ and MF$_3$ (the middle panel), are not correlated with TSI, since their CCs are insignificant (lie between $-$0.08 and 0.01).
To be more specific, the bottom panel in Figure~\ref{fig_xc} shows that the CCs between MF$_2$ and TSI are negative. The CC bottoms out at $-0.54$ at a time lag of around 0,
which means that the maximum of  TSI occurs at the time when the flux of MF$_2$ reaches its minimum. 
In the upper panel, it shows that \mfar\ and TSI are positively correlated, and their CC at lag 0 is 0.64. The CC peaks at 0.75 when \mfar\ is lagged with respect to TSI by 27 days, which means that the former probably leads the latter by a solar rotational period  on average in phase.
 The CC between MF$_4$ and TSI at lag 0 (0.76) is larger than that of \mfqr\ (0.70), and also larger than that of \mfar.

It also shows in the upper panel that the variation of the CC between \mfar\ and TSI around lag 0 behaves differently than that of the other two (MF$_4$ and \mfqr). In other words,  CC of \mfar\ shows a dip near lag 0, while both CCs of MF$_4$ and \mfqr\ show a peak. 
It means that, although the overall impact of  \mfar\ on  TSI is positive (enhancing TSI) at long-term timescales, there is a part in \mfar\ which should have an instant negative (reducing TSI) impact at timescale of the solar rotation period. This situation happens to neither MF$_4$ nor \mfqr. 
The waxing and waning of the CCs in all of the three panels also indicate short-term modulations that are related with the rotation of the Sun: the CCs show dips or peaks at lags that are integral multiples of the $\sim$27-day rotational period, i.e., -27, 0, 27, 54 days, etc. 
The correlograms show both long-term and short-term correlations between the data sets.

To confirm whether the lead of \mfar\ with respect to TSI is significant, and also to compare the modulations of MF$_4$, \mfqr\, and \mfar\ on TSI, we make use of the bootstrap method.
The bootstrap method is a resampling procedure that is suggested by Efron for the first time, and it has been widely used to handle statistical inference problems in many research areas. The primary advantage of this method is that it does not require a stringent model assumption on the underlying random process that generates the data. 
In this study, the non-overlapping block bootstrap method is used, and a detailed description of the specific process can be found in \citet{xu_phase_2017}.
Basically, each application of the non-overlapping block bootstrap method with cross-correlation analysis yields  five thousand pairs of CCs.
Note that the bootstrap method is used here for determining the uncertainties of the CCs between MF$_4$, \mfar\ and TSI,  but not for determining whether the MFs and TSI are significantly correlated or not. The physical causation between them (the surface magnetic field and TSI) is widely accepted (see, e.g., \citealt{domingo_solar_2009}).

\begin{figure}[tbp]
 	\centerline{\includegraphics[width=0.48\textwidth,clip=]{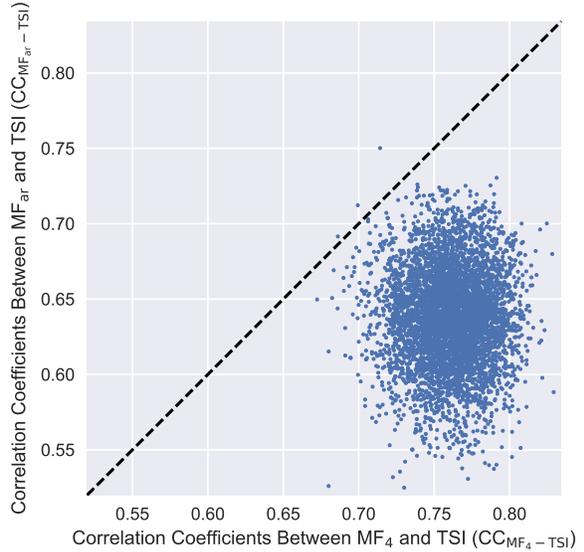}}
	\caption{Scatter plot of the correlation coefficients between MF$_4$ and TSI against that between \mfar\ and TSI. The dashed diagonal indicates the line where a point will fall in if the two coefficients are equal. \label{fig_sca}}
\end{figure}

\begin{figure}[tbp]
 	\centerline{\includegraphics[width=0.48\textwidth,clip=]{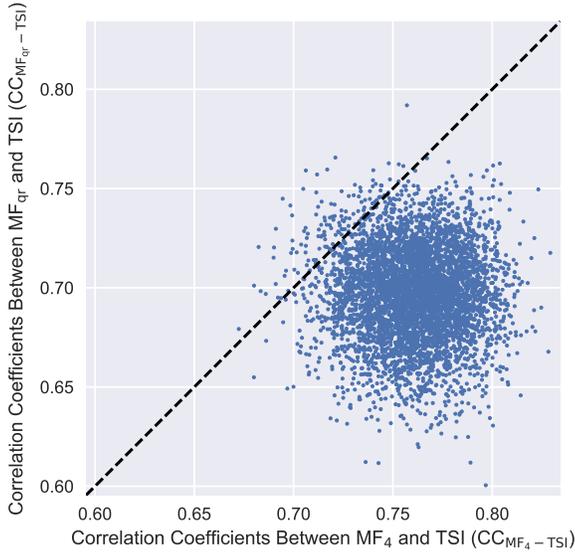}}
	\caption{The same as Figure~\ref{fig_sca} but between MF$_4$ and TSI against that between \mfqr\ and TSI. \label{fig_sca2}}
\end{figure}

To compare the correlations between MF$_4$, \mfar\ and TSI, we perform the bootstrap method on between MF$_4$, \mfar, and TSI, respectively, and 5000 CCs between MF$_4$ and TSI (CC$_\mathrm{MF_4-TSI}$) and 5000 CCs between \mfar\ and TSI (CC$_\mathrm{MF_{ar}-TSI}$) are obtained. They are plotted in Figure~\ref{fig_sca}. Figure~\ref{fig_sca} indicates that only 3 out of the 5000 points are located above the diagonal, which means that only 3 CC$_\mathrm{MF_4-TSI}$ are smaller than CC$_\mathrm{MF_{ar}-TSI}$.
Therefore, the probability that CC$_\mathrm{MF_4-TSI}$ is larger than CC$_\mathrm{MF_{ar}-TSI}$ is 99.94\%.
With the bootstrap method, 5000 CCs of \mfqr\ versus TSI (CC$_\mathrm{MF_{qr}-TSI}$) are also calculated. The CC$_\mathrm{MF_4-TSI}$ against CC$_\mathrm{MF_{qr}-TSI}$ are plotted in Figure~\ref{fig_sca2}. There are 137 points located above the diagonal, thus the probability that CC$_\mathrm{MF_4-TSI}$ is larger than CC$_\mathrm{MF_{qr}-TSI}$ is about 97.3\%. 

The bootstrapping tests confirm that CC$_\mathrm{MF_4-TSI}$ is larger than both CC$_\mathrm{MF_{qr}-TSI}$ and CC$_\mathrm{MF_{ar}-TSI}$, which infers that the impact of MF$_4$ on TSI is more significant than that of \mfar\ and \mfqr.

\begin{figure}[tbp]
 	\centerline{\includegraphics[width=0.48\textwidth,clip=]{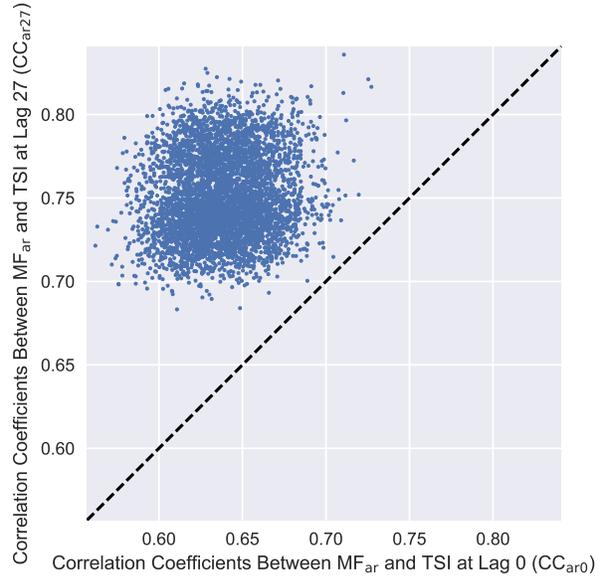}}
	\caption{Scatter plot of the correlation coefficients between \mfar\ and TSI at lag 0 against that at the lag of 27 days.
	\label{fig_sca27}}
\end{figure}

In order to confirm the phase relation between \mfar\ and TSI, another bootstrapping is carried out. 
5000 CCs between \mfar\ and TSI at lag 0 (CC$_\mathrm{ar0}$) and 5000 CCs between them at a 27-day lag (CC$_\mathrm{ar27}$) are obtained, and they are plotted in Figure~\ref{fig_sca27}.  
As can be seen in Figure~\ref{fig_sca27}, all of the points are located above the diagonal, which implies that the probability that CC$_\mathrm{ar27}$ is larger than CC$_\mathrm{ar0}$ is approximately 100\%.  Therefore, it means that \mfar\ indeed leads TSI by 27 days or about one rotational period.

At the timescale of a solar rotational period, TSI is modulated by the tradeoff of darkening by sunspots and brightening by faculae, together with the brightening of small-scale network elements in and outside active regions~\citep{hudson_effects_1982}.  
Faculae and sunspots usually occur at the same time in active regions. 
As shown in Figure 1 in~\cite{jin_suns_2011}, the active region areas defined are much larger than the area of sunspots seen in a continuum image, thus the flux in active regions \mfar\ includes flux from areas of both `dark' sunspots and `bright' faculae whose impacts on TSI are opposite.
 
The evolutionary behaviors of the sunspots and faculae are very different from each other.
The lifetime of faculae is much longer than that of sunspots: sunspots usually decay and disappear after one rotation, but faculae can live as long as several rotation periods~\citep{frohlich_total_2012}. 
When sunspots first appear on the photosphere, they cause an instant reduction on TSI especially in the visible wavelength band. 
However, despite the obvious darkening effect of sunspots, there are faculae and networks that cause even more significant brightening and overcompensate TSI, which explains why the cross-correlation between \mfar\ and TSI shows an overall positive relation with a sharp dip at lag 0.
After one rotation, sunspots usually decay into small-scale magnetic elements, but faculae are still alive, which leads to the increase of TSI. That is why TSI lags \mfar\ by a solar rotation period. 

To further investigate the modulations of the MFs on TSI, particularly at the rotational timescale, we decompose them and extract their rotational parts as in the following section.

\subsection{Complete Ensemble Empirical Mode Decomposition with Adaptive Noise (CEEMDAN)}

Empirical Mode Decomposition  (EMD) is an adaptive algorithm which decomposes a signal into a set of functions, known as Intrinsic Mode Functions (IMFs), and a residual called a trend. The IMFs are extracted from the signal itself using a process called sifting, and they represent the intrinsic   high- and low-frequency oscillations in the original signal. A complete description of this algorithm can be found in~\cite{huang_empirical_1998}. 

However, in the original EMD method, there is a problem named “mode mixing” which means that the presence of disparate frequency oscillations in the same mode, or the presence of similar oscillations in different modes~\citep{huang_review_2008}. 
To overcome the ``mode mixing" problem, \cite{wu_ensemble_2009} proposed an improved method named the Ensemble Empirical Mode Decomposition (EEMD). In this method, the EMD is performed over an ensemble of the signal plus Gaussian white noise. 
Nevertheless, it brings new problems: the reconstructed signal contains residual noise; and even more seriously, different realizations can produce different numbers of IMFs. To resolve this problem, a variation of EEMD is proposed by~\cite{m.e.torres_complete_2011}: the Complete Ensemble Empirical Mode Decomposition with Adaptive Noise (CEEMDAN). The CEEMDAN provides an exact reconstruction of the original signal and a better spectral separation of the modes, as well as advantages including less number of sifting iteration and lower computational cost.

The EMD (and EEMD, CEEMDAN) method does not require a priori assumption about the signal type, therefore it is widely and successfully used in the analysis of non-linear and non-stationary real-world signals, such as the variability of total solar irradiance~\citep{li_why_2012, lee_27-day_2015}, the periodicity of flare index~\citep{gao_periodicity_2011}, the research of solar mean magnetic field~\citep{xiang_ensemble_2016}, the sunspot numbers~\citep{barnhart_analysis_2011}, etc (either EMD or EEMD is used in these studies). 
This method is more effective than the traditional Fourier analysis in extracting modes with time-variable amplitudes and time-varying frequencies  in signals. In contrast, the latter method only decomposes data into fixed amplitudes and frequencies. Due to the fact that the Sun is not a rigid body and thus its rotation is a differential rotation with varying frequencies, the CEEMDAN is particularly suitable in this study.

\begin{figure*}
 	\centerline{\includegraphics[width=0.48\textwidth,clip=]{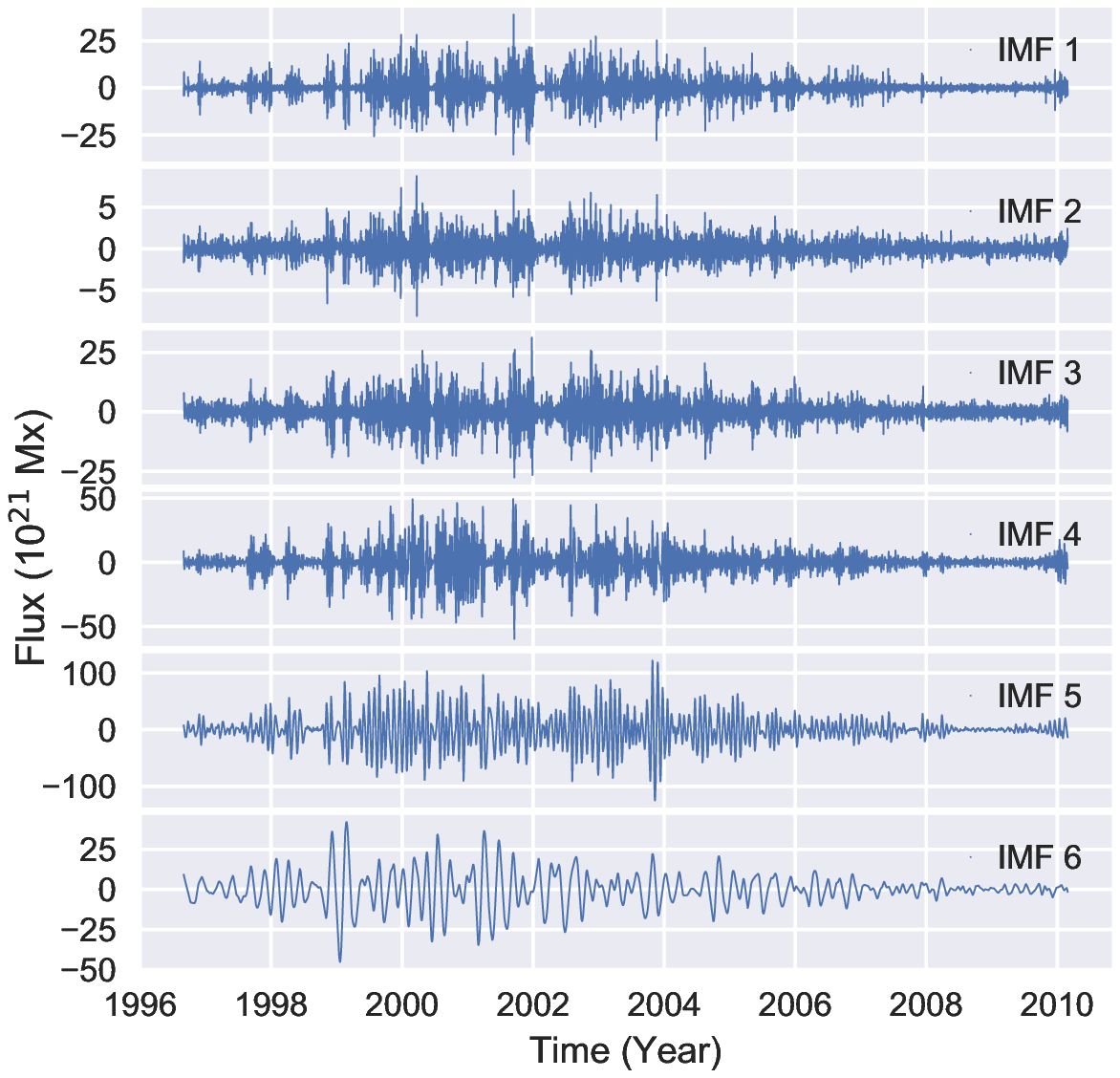}
	\includegraphics[width=0.48\textwidth,clip=]{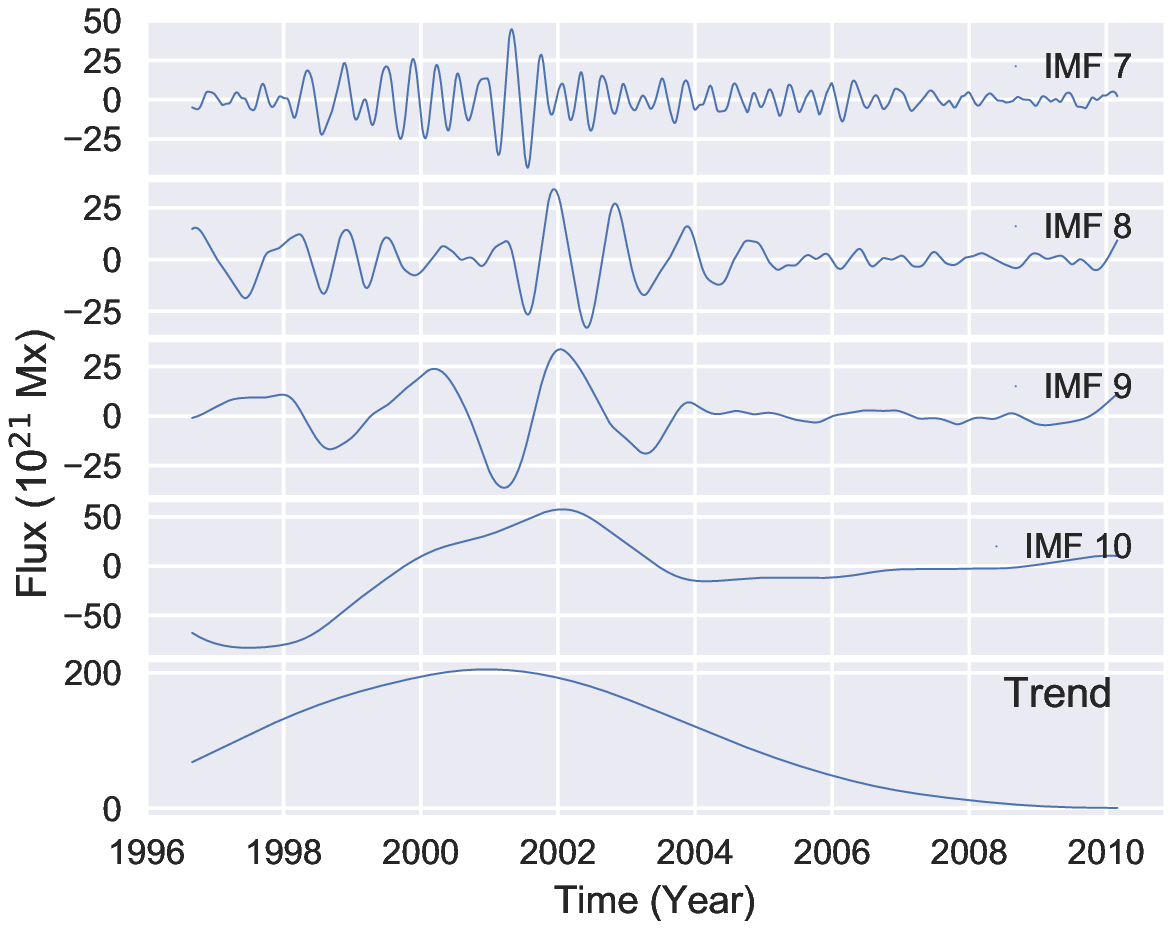}}
	\caption{The temporal variation of the intrinsic mode functions (IMFs) of  magnetic flux in active regions (\mfar) extracted by CEEMDAN. The last one in the right panel represents the trend. \label{imf}}
\end{figure*}

In order to determine and compare the variations in the MFs and TSI on variable timescales, we use the CEEMDAN to extract the temporal variations in them. The missing values in the data sets are interpolated with piecewise cubic Hermite interpolating polynomials.
The temporal variations of the IMFs of magnetic flux in active regions (\mfar) are shown, as an example, in Figure~\ref{imf}. It is obviously shown that the extracted IMFs range from high frequency (days) to low frequency (years). The last mode represents a nonlinear trend in \mfar, which indicates an estimate of the \mfar\ variation at the timescale of the $\sim$11-year sunspot activity cycle. The sum of these ten IMFs (IMF1--10) and the trend is identical  to the original data set of \mfar; that is to say, \mfar\ can be entirely represented by IMF1--10 together with the trend.

The 10 extracted IMFs are supposed to be modes in \mfar\ with different frequencies or periods, although the period in each mode may not be constant.
Then, for each IMF of \mfar, we  apply the Lomb-Scargle periodogram (LSP)  so as to obtain its period of oscillation. 
The LSP method is very useful in determining periodicities in time series even when there is a large amount of missing  samples, and its algorithm is introduced in detail in~\cite{xu_rotation_2016}.
The obtained periods of IMF3--9 of \mfar\ are show in the third row in Table~\ref{tbl-prd-emd}.

\begin{table*}
\caption{The Periods of the Intrinsic Mode Functions (IMFs) of MF$_2$, MF$_4$, \mfar\ and TSI, respectively (the unit is days). The IMFs and the trend of \mfar\ are shown in Figure~\ref{imf} as an example.}
\begin{center}
\begin{tabular}{lccccccc}
\hline\noalign{\smallskip}
Data set   & IMF 3 & IMF 4 & IMF 5 & IMF 6 & IMF 7 & IMF 8 & IMF 9 \\ 
\hline\noalign{\smallskip}
MF$_2$  & 6.6 & 13.4 & 26.2 & 59.6 & 176.7 & 357.4 & 970.2 \\
MF$_4$  & 7.2 & 13.4 & 26.2 & 59.1 & 177.3 & 370.5 & 814.3 \\
\mfar     & 6.9 & 13.2 & 26.6 & 65.5 & 112.4 & 319.5 & 778.3 \\
TSI          & 6.7 & 12.0 & 26.7 & 74.3 & 112.3 & 374.9 & 681.2  \\
\noalign{\smallskip}\hline
\end{tabular}
\end{center}
\label{tbl-prd-emd}
\end{table*}

Similarly, we applied the CEEMDAN and LSP analyses on MF$_2$, MF$_4$, and TSI, respectively. We obtained their IMFs, as well as their corresponding periods, which are also tabulated in  Table~\ref{tbl-prd-emd}.

\begin{figure}[tbp]
 	\centerline{\includegraphics[width=0.48\textwidth,clip=]{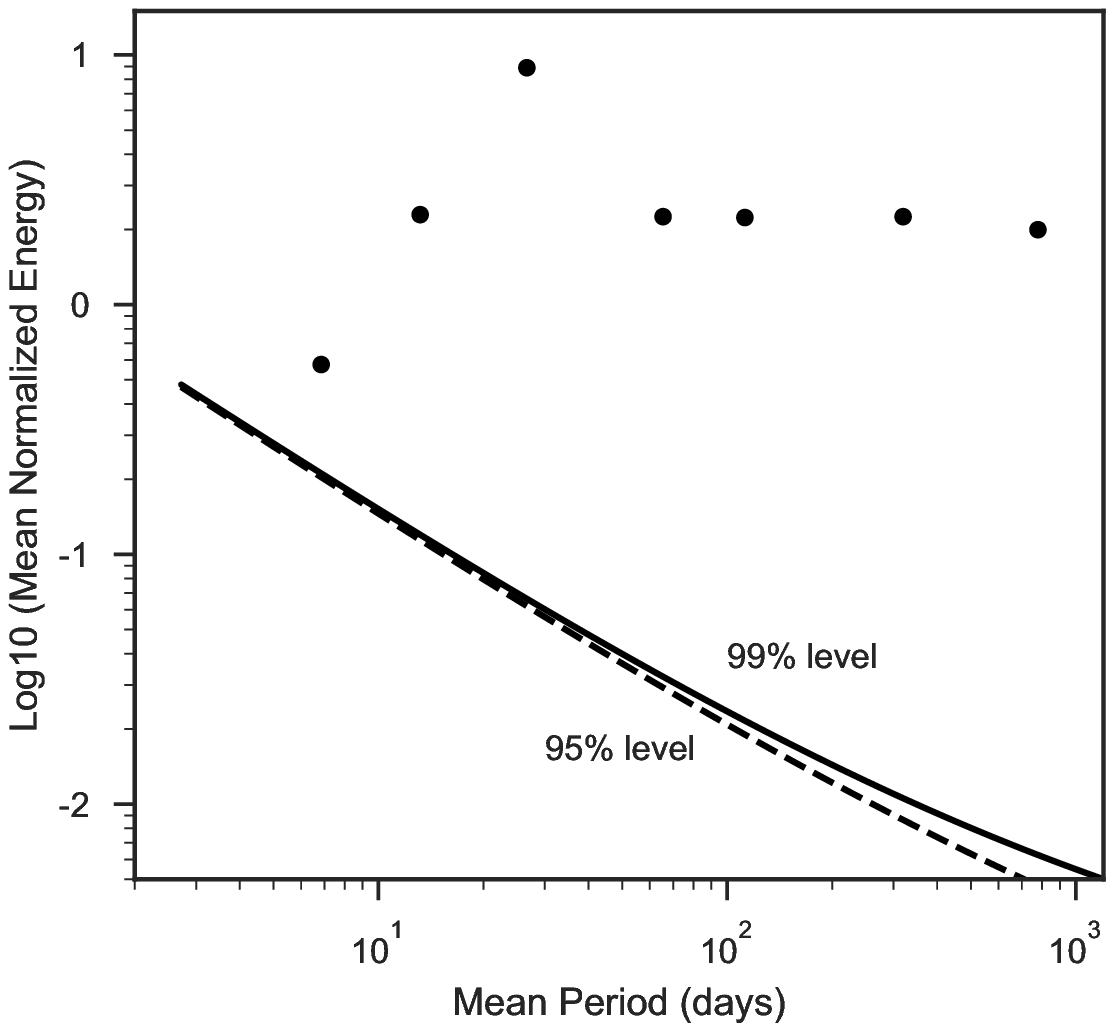}}
	\caption{Statistical significance test of the seven IMFs (IMF3--9) that are extracted from the \mfar. Each dot sign represents the Log10 (Mean Normalized Energy) of an IMF as a function of its Mean Period (days), ranging from the third IMF to the ninth IMF. The solid line represents the 99\% significance level and the dashed line is the 95\% significance level.
	\label{fig_sigTest}}
\end{figure}

It is found that the periods of the fifth modes of MF$_2$ and MF$_4$ are both 26.2 days, and that of \mfar\ and TSI are 26.6, 26.7 days, respectively, as indicated in the 4th column in Table~\ref{tbl-prd-emd}. Obviously, the fifth mode (IMF5) of both the MFs and TSI represent their respective $\sim$27-day rotational variations, and IMF4, IMF6 are inferred to represent their 13.5-, 54-day period variations.  
In order to ensure that an IMF contains a true signal, we test the statistical significance of the seven IMFs (Figure~\ref{fig_sigTest}). The statistical significance of the periods obtained can be determined by comparing spectral powers between the data set and white noise data~\citep{wu_study_2004}. When the white noise significance test is applied, all the 13.5-, 27-, 54-day periods of the IMFs are statistical significant within the 99\% significance level. 
\cite{xu_rotation_2016} investigated the rotational characteristics of the solar surface magnetic field with the same magnetic flux data sets. By applying the LSP to the original data sets of MFs, they found that the rotational period of MF$_2$, MF$_4$, and \mfar\ are 26.20, 26.23, and 26.66 days, respectively, which is in agreement with the periods that found here only with their IMF5s. It strongly suggests that their IMF5s should represent the rotational variations in them.  

\begin{figure*}
 	\centerline{\includegraphics[width=0.85\textwidth, trim=.5cm 1.5cm .5cm 2.cm, clip=]{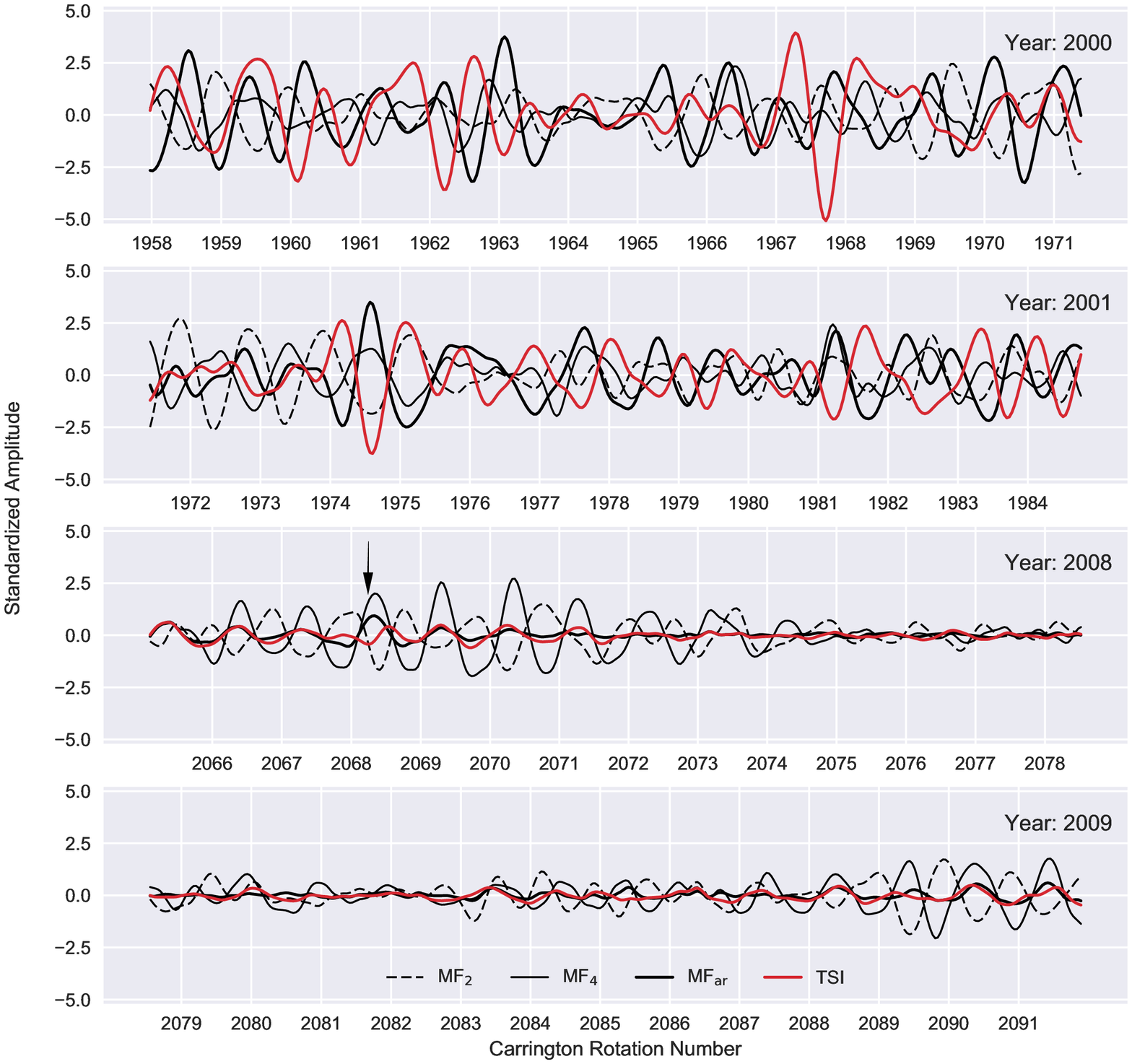}}
	\caption{Comparison of rotational variations of MF$_2$ (dashed), MF$_4$ (solid), \mfar\ (thick solid), and TSI (red) during four years. The years 2000, 2001 (upper two panels) are selected to represent solar maximum time, and the years 2008, 2009 (bottom two panels) are selected to represent solar minimum time. The abscissa denotes Carrington rotational cycle, and the vertical coordinates indicate the standardized amplitude of the rotational variations of both MFs and TSI. Black arrow in the year 2008 (Carrington rotation number 2068) indicates the time of a 2008 outburst, when the TSI and MFs variations are dominated by a single active regions. Note that the range of y-axis in the four panels are all the same.
	\label{min_maxY}}
\end{figure*}

To further study the amplitude and phase of the $\sim$27-day variations of MFs and their modulations to those of TSI, the IMF5s of the four data sets are shown in Figure~\ref{min_maxY}.  The years 2000 and 2001 (the upper two panels) are shown as representatives of the solar maximum time, and the years 2008 and 2009 (the bottom two panels) are shown as representatives of the solar minimum time. Note that the abscissa denotes the Carrington rotation number (CRN). In order to compare them clearly, all of the IMF5s are standardized. For example, the standardized amplitude of the IMF5 of \mfar\  (as shown in the left panel of Figure~\ref{imf}) is estimated by removing its mean and then dividing with its standard deviation: $\mathrm{A_{IMF5} = \frac{IMF5(t)-\overline{IMF5}}{\sigma(IMF5)}}$.

As can be seen in Figure~\ref{min_maxY}, during solar maximum time, the relative amplitudes of the MFs (MF$_2$, MF$_4$, and \mfar) are all larger than that during solar minimum time; the situation of TSI is the same. 
Besides, all of them display rotational variations, i.e., waxes and wanes with an approximate $\sim$27-day period.
During the time when there are no (or very few) active regions, such as from CRN2074 to 2086, all of the relative amplitudes are very small. During solar maximum time, the variations of MF$_4$ and \mfar\ are obviously out-of-phase with that of TSI; in contrast, the former two seems inphase with the latter during solar minimum time.

The relative amplitudes and phase relations of the MFs and TSI are related with the emergence and evolution of active regions on the rotating solar photosphere. 
Magnetic flux tubes in the subsurface region emerge to the solar photosphere and form active regions very quickly, and the process lasts for up to 5 days. However, the dispersion of an active region takes much longer. Depending on the property of the active region, it takes 4 to 10 months for an active region to disperse to the magnetic flux density level of the quite-Sun~\citep{cheung_life_2016}. 
As for sunspots, according to the Gnevyshev-Waldmeier rule, most of the sunspots only live for less than a day~\citep{solanki_sunspots:_2003}.
At solar maximum time as in the years 2000 and 2001, active regions emerge to the surface frequently. Therefore, TSI is constantly modulated by a compound effect of multiple active regions. So the quantities of flux in the MFs vary quickly, and the relative amplitudes of the MFs and TSI are larger. Since sunspots have a negative instant impact on TSI, \mfar\ and TSI are not inphase or even in antiphase during most of time at solar maximum as shown in the upper two panels.

In 2008 and 2009, there are 516 out of the 730 days (72.3\%) when there were no sunspots  (sunspot number equals to 0) on the solar surface. The minimum of solar cycle 24 is in a very low level of magnetic activity~\citep{li_investigation_2016}.
During solar minimum time, active regions rarely appear, therefore the modulation of a sole active region can be clearly seen. For example, there is an outburst in CRN2068 in 2008 as indicated with a black arrow in the third panel of Figure~\ref{min_maxY}.  It can be seen that the amplitudes of \mfar, MF$_2$, MF$_4$ and TSI all increased after the burst. The amplitude of \mfar\ and TSI first increased and then decline to tranquility after about 4 rotation periods until CRN 2072. \mfar\ and TSI are in antiphase during the following around ten days, and then become inphase when the sunspot disappears and TSI is only modulated by `bright' faculae and network features. The outburst can be seen obviously from the variations of the MFs and TSI here; it can also be identified easily with the sudden rise of the sunspot numbers.
Note that it is due to the outburst in late 2007 that the amplitudes of the MFs and TSI at the beginning of 2008 (before CRN2067) are also relatively large. 

\begin{figure}[tbp]
 	\centerline{\includegraphics[width=0.48\textwidth, clip=]{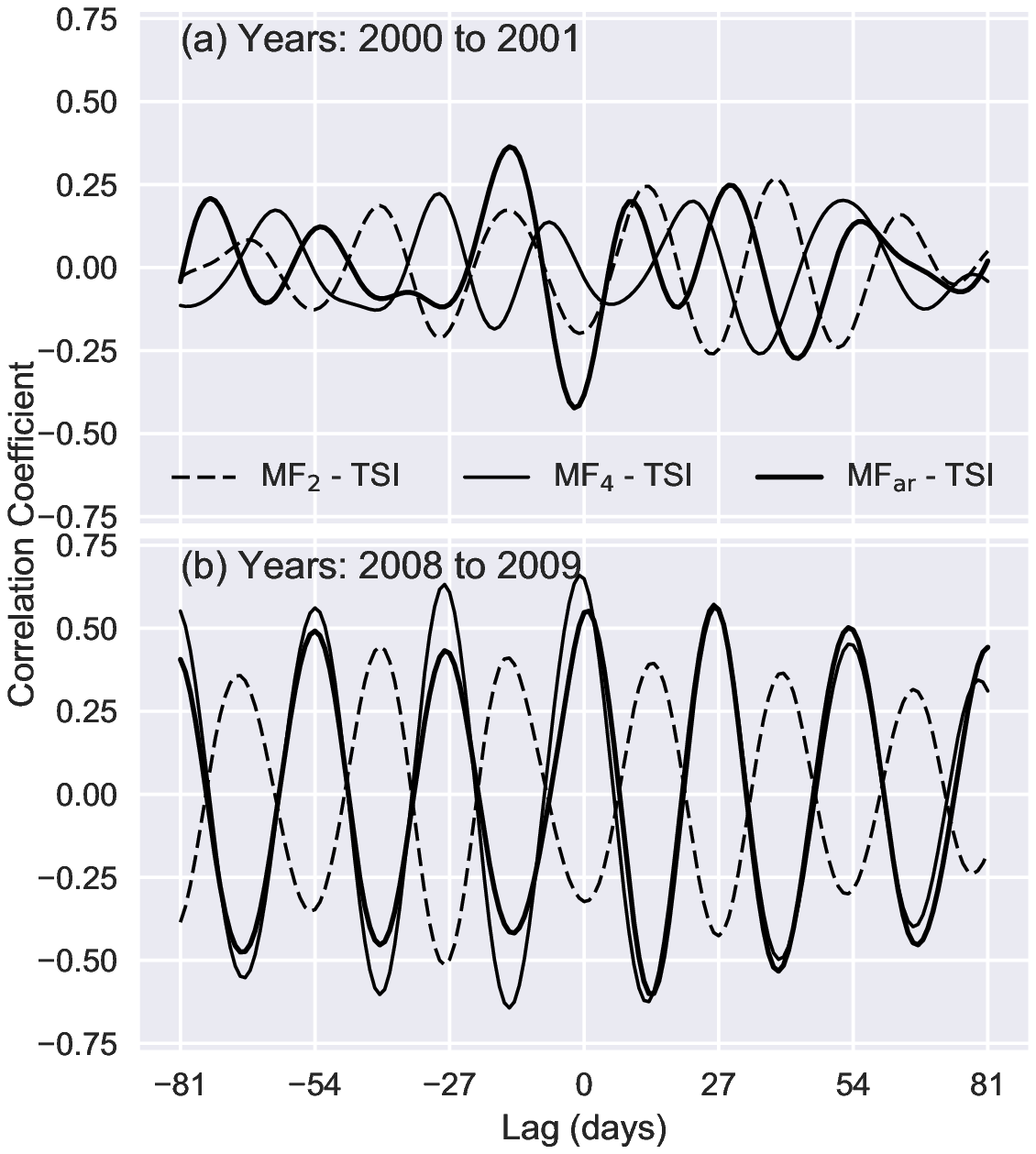}}
	\caption{Cross-correlograms  between the rotational variations of MF$_2$ (dashed), MF$_4$ (thin), \mfar\ (thick), and TSI during the maximum (years 2000 to 2001) and minimum (years 2008 to 2009) times. The rotational variations, i.e., IMF5, of the four years are shown in Figure~\ref{min_maxY}.
	\label{xc_shortterm}}
\end{figure}

In order to show the phase relations between the rotational variations of the MFs and TSI during  maximum (years 2000 to 2001) and minimum (years 2008 to 2009) times more clearly, a cross-correlation analysis is performed  between the IMF5s of MF$_2$, MF$_4$, \mfar, and TSI, respectively, and the result is shown in Figure~\ref{xc_shortterm}. 
It is depicted in the upper and lower panels that \mfar\ is in antiphase with TSI  during solar maximum and inphase during solar minimum, while MF$_2$ always shows an antiphase relation with TSI. It also implies that the impact of MF$_2$ on TSI is similar at different timescales (referring to the bottom panel in Figure~\ref{fig_xc}). 
As for MF$_4$, it is inphase with TSI during solar minimum; and during solar maximum, there is a phase difference between them.
The variations of all of the CCs indicate a clear regular rotational modulation. 

At solar maximum times, TSI is usually continuously modulated by multiple sunspots (and active regions where they are situated) on the solar surface at the same time. Due to the instant darkening effect of sunspots, \mfar\ and TSI are anti-correlated at the solar rotational timescale, as shown in the upper panel.
During solar minimum time, there can be only faculae but no sunspots; even when there are small sunspots emerge occasionally, they decay and disappear soon. Thus, the TSI is then mainly solely modulated by faculae and networks which cause the positive correlation between \mfar\ and TSI as shown in the lower panel.

\begin{figure}[tbp]
 	\centerline{\includegraphics[width=0.48\textwidth, clip=]{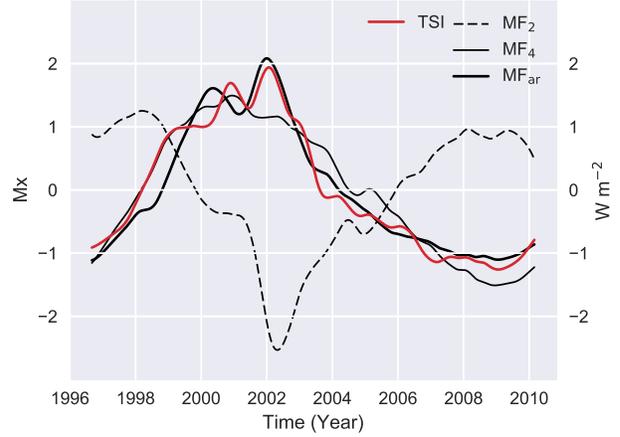}}
	\caption{Comparison of the relative amplitudes of long-term variations with longer than 500-day periods from MF$_2$ (dashed, left y-axis), MF$_4$ (solid, left y-axis), \mfar\ (thick solid, left y-axis), and TSI (red, right y-axis). The ranges of the left and right y-axis are the same. \label{low_freq}}
\end{figure}

With the CEEMDAN, both TSI and the MFs can be separated into short-term variations with shorter than around one year period, and long-term variations with longer than one year period. The one year period, which is approximately the periods of IMF8 of the MFs and TSI, is chosen because that is about the uppermost time for an active region to decay and disperse into quiet Sun network magnetic field~\citep{cheung_life_2016}.
The relative amplitudes of the long-term variations of the MFs and TSI are shown in Figure~\ref{low_freq}.  

It is displayed in the figure that the intra-cycle long-term variations of MF$_4$ and \mfar\ are in very good agreement with that of TSI, while MF$_2$ is in obvious antiphase with them.

\begin{figure}[tbp]
 	\centerline{\includegraphics[width=0.48\textwidth, clip=]{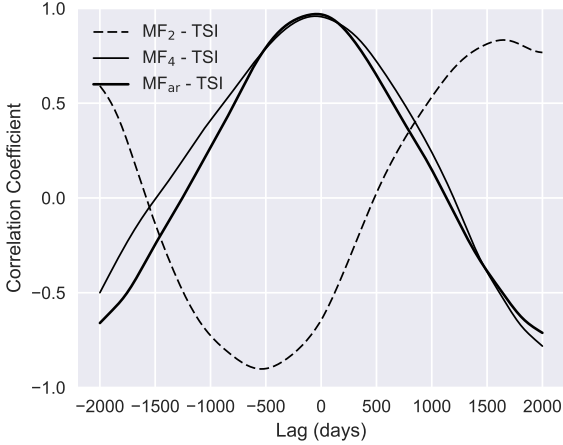}}
	\caption{Cross-correlograms between long-term variations of  MF$_2$ (dashed), MF$_4$ (thin), \mfar\ (thick), and TSI that are shown in Figure~\ref{low_freq}.
	\label{xc_longterm}}
\end{figure}

To further confirm their phase relation, another cross-correlation analysis is performed and the result is shown in Figure~\ref{xc_longterm}. The correlation functions of MF$_4$ and TSI and that of \mfar\ and TSI are very similar, and they peak at around lag 0 with a coefficient of 0.96 and 0.97, respectively. The CC between MF$_2$ and TSI at lag 0 is $-0.65$, and it bottoms out at $-0.90$ at a 539-day lag, which is around 1.5 years; it implies that the long-term part of MF$_2$ lags (negatively) that of TSI by around one and a half year, and the reason for this is not clear. 
\cite{xu_rotation_2016} speculated that the magnetic elements contributing to MF$_2$ may mainly come from a higher latitude or emerge from a subsurface shear layer named the leptocline.

\section{Conclusion} 

There is increasing evidence that the variations of solar irradiance is one of the key factors that  influence the global climate. The surface magnetic field plays an important role in the variability of the solar total and spectral irradiance~\citep{domingo_solar_2009}.
Using daily values of solar photospheric magnetic flux data sets at four specific ranges (MF$_{1\sim4}$) together with the total flux of active regions (\mfar)  and quiet regions (\mfqr), we examined and characterized the intra-cycle modulation of different magnetic flux to the variability of TSI. The time interval of the data sets is from 1996 September 1 to 2010 February 28.
 The data sets used in this study is unique due to the fact that it is the only space-borne measurements of the full-disk solar magnetic field with a consistent sensitivity and resolution for a cycle-long time interval.

Cross-correlation analyses between the full data sets of MF and TSI show that MF$_2$ is negatively correlated with TSI, while MF$_4$, \mfar, and \mfqr\ are all positively correlated with TSI. In addition, both MF$_1$ and MF$_3$ show insignificant correlation with TSI. The variations of the correlation coefficients indicate obvious rotational modulations.

Bootstrapping tests confirm that the correlation coefficient between MF$_4$ and TSI is larger than that of both \mfar\ and \mfqr, which implies that the modulation of magnetic elements in range MF$_4$ on the variations of TSI is more significant than that of active regions and quiet regions. The sharp dip at lag 0 of the CC between \mfar\ and TSI indicates that part of the flux in the former should have an instant negative (reducing) impact on TSI despite the overall positive (enhancing) impact, we infer that it is due to the darkening effect of sunspots.
 The peaks at around lag 0 of the CC between MF$_4$, \mfqr\ and TSI show their positive instant impacts on TSI. 
The cross-correlogram and bootstrapping between \mfar\ and TSI display that the former should lag the latter by 27 days, or around one solar rotation period. This is due to the differences of the effect and lifetime of sunspots and faculae which are the main magnetic features that modulate TSI on the rotational timescale.

By extracting the rotational variations in the MFs and TSI with the CEEMDAN method, the modulations of the former on the latter at the solar rotational timescale are clearly shown and compared during solar maximum and minimum times, respectively. 
At solar maximum times, there are usually multiple active regions exist on the solar surface at the same time, thus the relative amplitudes of the MFs and TSI are all much larger compared with that at solar minimum time. \mfar\ is negatively correlated with TSI, which is a result of the instant darkening effect of multiple sunspots. 
The correlogram during minimum time shows that both MF$_4$ and \mfar\ are inphase with TSI. 
The correlation between MF$_2$ and TSI is negative at either solar maximum or solar minimum, and the impact of the former on the latter is similar at different timescales. 
All of the CCs show a clear variation with the solar rotational period.
The modulation of an unique sunspot can be seen clearly at solar minimum, which is illustrated in detail with the 2008 outburst in CRN2068. 

Comparison of the relative amplitudes of the long-term variation shows that TSI is in good agreement with the variation of  MF$_4$ and \mfar; their CCs are as high as 0.96 (between MF$_4$ and TSI) and 0.97 (between \mfar\ and TSI) around 0 lag. MF$_2$ is in antiphase with them, and it lags TSI by about 1.5 years.


\acknowledgments
The authors thank C. L. Jin very much for providing the data. 
The authors are grateful to Prof. K. J. Li for his valuable suggestions and discussions.
This work has made use of open-source softwares including Numpy~\citep{walt_numpy_2011}, Scipy~\citep{jones_scipy:_2001}, Matplotlib~\citep{hunter_matplotlib:_2007}, IPython~\citep{perez_ipython:_2007}, and Jupyter.\footnote{\url{https://jupyter.org}}
This work is supported by the National Natural Science Foundation of China (11573065, 11633008, 11273057 and 11603071), the Specialized Research Fund for State Key Laboratories, and the Chinese Academy of Sciences.

\nocite{*}
\bibliographystyle{aasjournal}  
\bibliography{ref_flux_tsi}

\end{document}